# Gated MoS$_2$/SiN Nanochannel for Tunable Ion Transport and Protein Translocation

Shukun Weng [1,2,*]; Ali Douaki [1,3]; Makusu Tsutsui [4], German Lanzavecchia [1,3], Anastasiia Sapunova [1,2], Lorenzo Iannetti[5], Alberto Giacomello[5], Roman Krahne [1], and Denis Garoli [1,3*]

[1] Optoelectronics Research Line, Instituto Italiano di Tecnologia, 16163 Genova, Italy

[2] Department of Materials Science, University of Milano-Bicocca, Via R. Cozzi 55, I-20125 Milano, Italy

[3] Dipartimento di Scienze e Metodi dell'Ingegneria, Universitàdegli Studi di Modena e Reggio Emilia, Via Amendola, 2, 43122 Reggio Emilia, Italy

[4] The Institute of Scientific and Industrial Research, Osaka University, Mihogaoka 8-1, Ibaraki, Osaka, 567-0047, Japan

[5] Dipartimento di Ingegneria Meccanica e Aerospaziale, Sapienza Università di Roma, Via Eudossiana, 18, 00184 Roma, Italy

**Abstract**

Ionic transport in nanofluidic channels holds great promise for applications such as single-molecule analysis, molecular manipulation, and energy harvesting. However, achieving precise control over ion transport remains a major challenge. In this work, we introduce a MoS$_2$/SiN hybrid nanochannel architecture that enables electrical tuning of ionic transport via external gating, and we examine its potential for osmotic power generation and single-molecule detection. To fabricate the channels, we employed a combined focused ion beam (FIB) milling and dry transfer method, producing sub-10 nm thick structures while preserving the structural integrity and electronic properties of $MoS_2$—essential for reliable surface charge modulation. We first investigated how the gate voltage influences ionic conductance, finding evidence of gate-dependent modulation of ion selectivity under different bias polarities. Next, by applying a salt concentration gradient across the nanochannels, we demonstrated the feasibility of this platform for osmotic energy harvesting. Finally, we tested the system for single-molecule sensing, showing that linearized bovine serum albumin (BSA) produced translocation signals with notably long dwell times. Together, these results highlight gated $MoS_2$/SiN nanochannels as a promising platform for tunable nanofluidics, with potential applications in controlled molecular transport and energy harvesting from osmotic gradients.

**Keywords**: Solid-state Nanopores, 2D material, Nanochannel, Ion transport, single protein detection, osmotic power.

INTRODUCTION

Ion transport through nanochannels plays a crucial role in a wide range of applications, including drug delivery, biosensing, desalination, and energy harvesting. Among the various systems studied, nanochannels composed of two-dimensional (2D) materials such as $MoS_2$ and graphene have recently received significant attention. Their atomic thickness allows the creation of sub-nanometer channels, which provide a powerful platform for exploring fundamental phenomena such as quantum friction, ion adhesion, orientational conduction, and DNA translocation [1–4]. Despite these advantages, the fabrication of multi-stacked 2D material structures remains a challenging task, and the development of simpler device architectures would greatly facilitate the study of nanofluidic processes in this unique family of devices.

The strong confinement effect characteristic of 2D-material-based nanochannels results in distinctive ion transport properties. These properties can be modulated through adjustments in interlayer spacing, surface chemistry, or material composition [5,6]. In the context of single-molecule sensing, $MoS_2$ offers particular advantages compared with graphene. Its hydrophilic surface reduces unwanted interactions with DNA [4,7,8], while intrinsic sulfur vacancies on the $MoS_2$ surface provide sites that can selectively slow down or capture proteins containing free thiol groups. This capability highlights $MoS_2$'s potential in advanced protein sensing applications [9,10].

Among the many nanofluidic device concepts, one of the most intriguing is the ionic transistor, which is considered a vital component for future ion-based computing technologies. A key feature of such devices is the ability to electrically tune the surface charge of individual nanopores or nanochannels, thereby enabling precise control over ion transport [11–21]. In this regard, $MoS_2$ is particularly promising because its carrier type can be highly tuned by either surface modification or electrical gating [22–27]. By applying electrical gating, the surface charge of $MoS_2$/SiN nanochannels can be modified, allowing researchers to manipulate their ion transport behavior and optimize performance for specific applications. This electrical tunability not only deepens our understanding of the fundamental mechanisms underlying ion translocation but also opens practical opportunities in biosensing, molecular separation, and energy storage [28].

In this paper, we report a detailed investigation into the ion transport properties of $MoS_2$/SiN nanochannels, with a particular focus on their electrical tunability through gating. We also explore their potential applications in protein detection and electro-osmotic power generation. Our study examines the influence of ions, pH, and electrical gating on the surface charge of $MoS_2$ and analyzes how these factors affect ion transport. Through these experiments, we elucidate possible mechanisms of ion translocation in MoS$_2$/SiN nanochannels and discuss their implications for the design of nanofluidic devices. Overall, our findings contribute to the expanding knowledge of nanoscale ion transport and highlight the promise of electrically tunable MoS$_2$/SiN nanochannels as a platform for future advances in nanotechnology and biomedicine

EXPERIMENTAL METHODS

**Silicon Nitride (SiN) Membrane Fabrication**

Freestanding Si$_3$N$_4$ membrane chips were fabricated using a standard procedure. An array of square membranes was created on a commercial double-sided 100 nm LPCVD SiN-coated 500 μm Si wafer through UV photolithography, reactive ion etching, and KOH wet etching [29]. Depending on the dimension of the photolithography mask we can get different sizes of SiN membrane. Here we mainly used ~50*50 μm membrane to increase the robustness of our device and reduce the noise level [30].

**Nanostructure Fabrication**

Here, we propose a novel structure which can be fabricated by a simpler process compared with previously reported sandwiched 2D material stacks structures [31]. The nanopore and the nano-slit(s) are directly milled by focused ion beam (FIB) on suspended SiN membrane with 30 kV acceleration voltage and 18 pA current, after which a thick layer of MoS$_2$ is transferred on top of the nano-slit to form a nanochannel comprised by SiN and MoS$_2$ surface, as shown in Fig. 1a. The thickness of the MoS$_2$ flake is typically thicker than 50 nm to avoid significant sagging of the flake that can block the channel. The depth of the nanochannel is mainly controlled by the dwell time of the ion beam while milling the nano-slit, in our cases it's typically between 5 nm to 20 nm.

**2D Material Transfer**

There are mainly two ways to prepare MoS$_2$ flakes, physically by means of mechanical exfoliation (cleave the original crystal) or chemically (controlled chemical reaction on substrate by CVD/CVT). These two techniques result in different surface charge characteristics [32–34]. On the fresh cleaved MoS$_2$ flakes, the surface is near intrinsic because of the smooth surface, and the electrons will gradually accumulate on the surface because of the slowly desulfurization in the air, which induces a negatively charged MoS$_2$ surface.

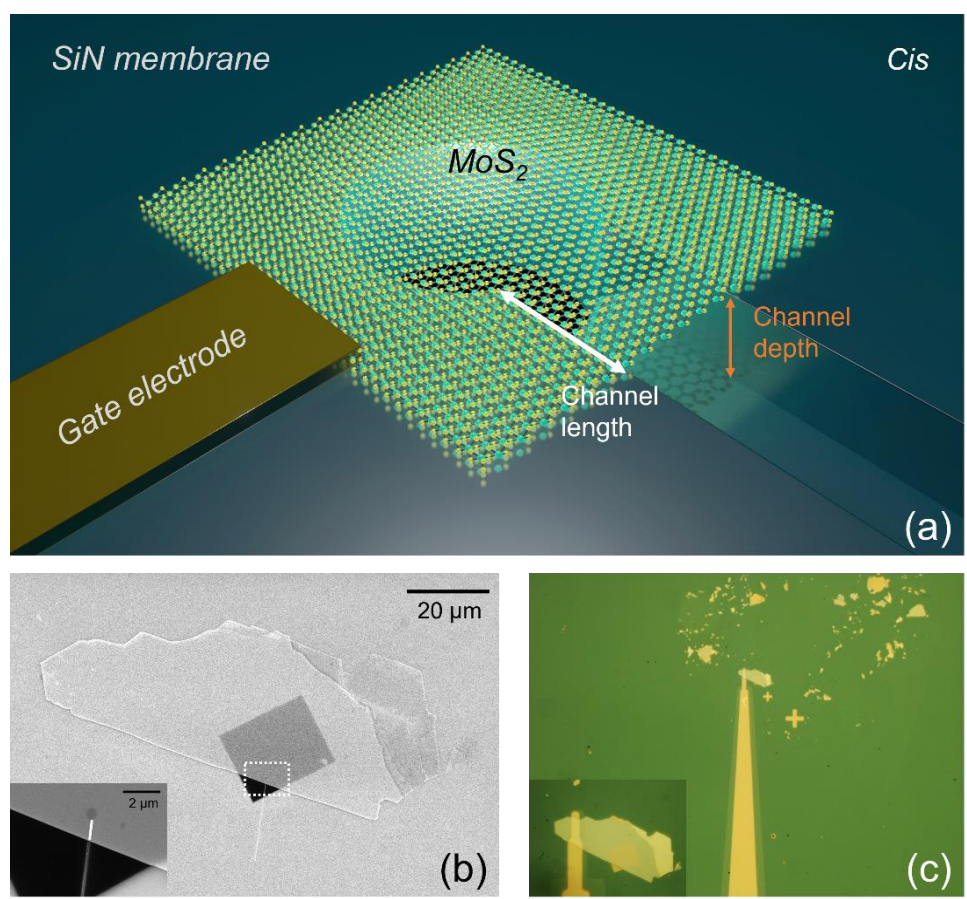

**Figure 1**. Device design. (a) Illustration of nanochannel devices formed by MoS$_2$ flake and SiN membrane with a nanopore and nano-slit. The length, width, and height of the nanochannel is around 1μm, 100 nm, and 10 nm respectively; (b) SEM image of nanochannel device just after the transfer of MoS$_2$; the inset shows the magnification image of white dash line framed region, and the solid white line in the inset image; (c) Optical image of nanochannel device after the Au electrode deposition.

On the other hand, the CVD/CVT grown MoS$_2$ flakes are always negatively charged because of the dangling bonds generate during the chemical reaction process [34]. MoS$_2$ flakes prepared via chemical processes can be suspended in proper solvents and easily deposited on solid-state substrates [35] but they typically present a high level of contamination making them not suitable for the scope of this work. Therefore, here, we used exfoliated MoS$_2$ flakes as top layer to form the nanochannels (Fig. 1b). The MoS$_2$ flakes were exfoliated using well established scotch tape methods onto a 300 nm SiO$_2$ substrate. By using a standard optical microscope it is possible to quickly obtain a first check on the quality of the transferred flakes. A polycarbonate (PC) thin film on top of PDMS dome supported by a transparent cover slide is used to transfer the selected MoS$_2$ flake from SiO$_2$ substrate to SiN membrane which has a nanopore connect with a shallow nano-slit (Fig. 1a) [36,37]. The PC thin film was prepared by dip-coating method. First a PC solution was prepared by dissolving polycarbonate in chloroform (10% weight ratio). The solution was then deposited onto a clean glass slide using a drop-casting method. A second glass slide was immediately positioned parallel above the coated slide and gently pressed with finger pressure to eliminate air bubbles between the two surfaces. Subsequently, the upper slide was rapidly removed through a horizontal sliding motion, leaving a thin, uniform layer of PC solution on the lower substrate. The chloroform solvent was allowed to evaporate under ambient conditions for 1 minute, resulting in the formation of an ultrathin, uniform PC film. This free-standing PC film, when combined with perforated double-sided adhesive tape mounted on a PDMS stamp, served as an effective transfer medium for 2D materials in subsequent processing steps [38].

**Gating electrode fabrication**

The Ti/Au electrode (Fig. 1) was fabricated using two photolithography steps. First, Au electrode region pattern was defined using the resists LOR3B/S1805 spin coated as a double layer. A 5 nm Ti adhesion layer followed by a 30 nm Au layer were deposited on developed region by using electron beam evaporation. followed by lift-off in PG remover for 5 min.

A second lithography process using a larger line-width pattern to isolate the Au electrode was performed using a similar process (spin coating photoresist, mask-aligner exposure, evaporation of isolated material and lift-off). As shown in Fig. 1c, the tip of the Au electrode contacts with the $MoS_2$ flake, while the pad is positioned outside the membrane for electrical characterization of ionic transistor behavior.

RESULTS & DISCUSSION

**Basic ion transport properties**

The first characterization of the platform was done by measuring the ionic current with different concentrations of KCl. The conductance of the nanochannel did not increase linearly when the concentration of KCl electrolyte increased from 1 mM to 1000 mM. This non-linear behavior could be attributed to mechanisms such as ionic coulomb blockade as described for sub-nm monolayer $MoS_2$ nanopore or the predominance of counterions transport [39,40]. While the ion current is very small due to the limited channel height and because of potential hydrocarbon contamination, we observed an ionic rectification with a rectification ratio up to 10 for 1M KCl as shown in Fig. S1.

In order to study the surface charge effect of the nanochannel toward ion current, the ion current under different pH of KCl were measured. In this case, we first applied a potential ranging from –100 mV to 100 mV with an alternating current scan overnight. The alternating electric field prompted ions to accumulate at the nanochannel entrance and repeatedly penetrate the contaminant barrier. Through multiple cycles, the blockages were gradually removed, thereby restoring the ion transport capacity of the channel(Fig. S2) [41]. Fig. 2(b) shows how the ion current is significantly affected by the pH value of the electrolyte after the overnight cleaning process. At lower pH values, the current is higher, likely due to the increased proton concentration.

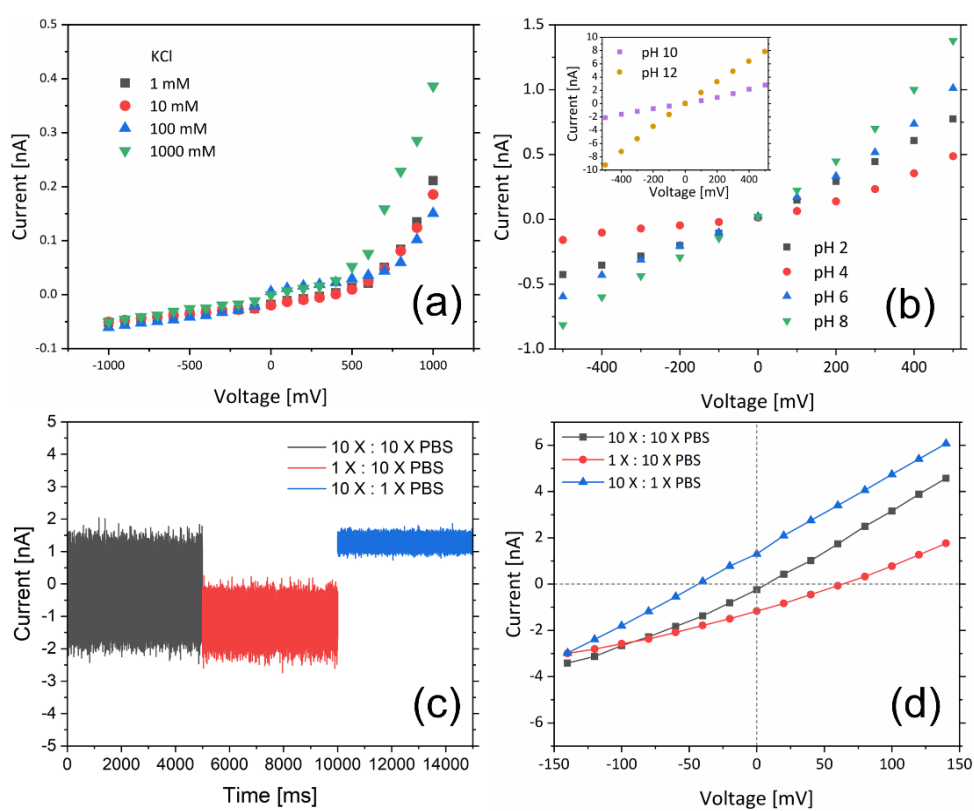

**Figure 2**. Ion transport of nanochannel. (a) Ion current across nanochannel with different KCl concentrations under pH 7.4 and (b) Ion current of 10 mM KCl across nanochannel with different pH; (c,d) Ion current of PBS buffer solution across nanochannel with different concentration in two side of membrane.

As the pH increases, the current increases, suggesting that the nanochannel surface becomes more negatively charged, attracting the positively charged ions. The minimal ion current value occurred at pH 4 attributable caused by the effective isoelectric point of the MoS$_2$/SiN channel, consistent with published isoelectric point data of MoS$_2$ and SiN [42,43]. With a pH increased from 10 to 12, the rectification direction changed. The pH-dependent behavior highlights the importance of surface charge in controlling ion transport through the nanochannel [44].

The ion rectification further shed light on energy harvesting from salinity gradients. Several recent studies, in fact, demonstrated how it is possible to generate energy from the osmotic process in nanopores / nanochannels using the buffers at the cis/trans sides with two significantly different concentrations [11,45–47]. Here, we first measured the response of our platform using 10X PBS buffer solution (1.37 M NaCl) in both side reservoir (Cis/Trans) obtaining almost zero ion current without applying any external voltage. On the contrary, replacing the electrolyte in the trans reservoirs to 1X PBS (0.137 M NaCl) we immediately got ion current around 1 nA. We also observed a higher noise level for 1:10X PBS (10X in Cis chamber) case with respect to the symmetrical case 10:1X PBS (10X in Trans chamber) as shown in Fig. 2c. This can be explained by the ion selectivity caused by the uneven distributed surface charges induced higher ion perturbation of one side of nanochannel entrance compared to the other side, which introduce high 1/f and white noise to the ion current as shown in Fig. S3 [48]. By scanning the external potential we can get IV curve from different configurations as shown in Fig. 2(d) where we can calculate the power of our osmotic generator ~230 pW by multiply the interception value of IV curve ($P = V_{osm}^2 G$, while $V_{osm} = V_{inteception} - V_{redox}$ is the effective osmotic potential and G is the conductance [49]), the redox voltage $V_{redox}$ set here as -41 mV was take from previous reference [47]. Noteworthy, the obtained osmotic power is higher than the value from polyimide nanochannel (26 pW) and monolayer MoS$_2$ nanopore osmotic generator (160 pW) [50,49]. The power density of our nanochannel can also be derived by considering the power per unit channel surface of the cross section $P_{density} = P_{osm}/A$, while the area of the cross section A is 1000 nm$^2$ so the power density of the nanochannel can be calculated as 230 kW/m$^2$ which is higher than that of hBN nanotube (~3000 W/m$^2$) [51]. In order to explore a potential scale-up of this system, we prepared a sample with 20 nanochannels distributed around a single window as depicted in Fig. S4. By characterizing this configuration using the same salt gradient, we obtained a power of about 760 pW, with a 3-folds enhancement with respect to the single channel. The non-linear enhancement with the number of nanochannels is most likely caused by variations in their size—particularly the channels' length and depth—which resulted in an overall lower current.

**Electrically Tunable Ion Transport**

Given that the ion current rectification (ICR) observed in the MoS$_2$/SiN nanochannel may result from an unbalanced surface charge distribution on the two channel walls, we investigated ion transport behavior as a function of MoS$_2$ surface charge by applying a potential through a top gate electrode. Fig. 3(a) shows the electrical measurement setup for the transistor-like configuration, using two channels from a Keithley 2612B source meter. As shown in Fig. 3(b), the channel conductance increases with decreasing gate voltage down to -500 mV, which can be attributed to a higher counter-cation concentration induced by the enhanced negative surface charge on the nanochannel walls.

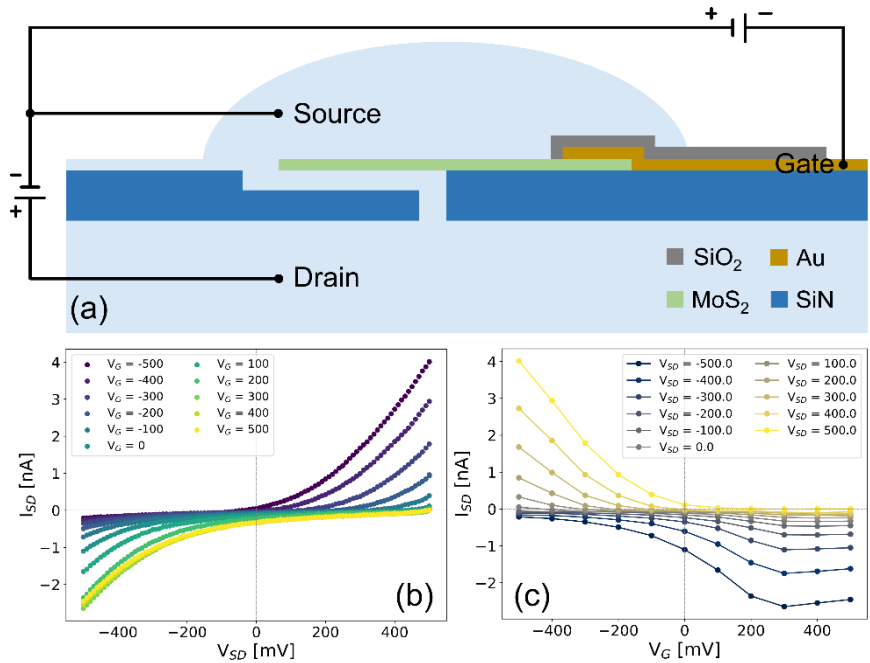

**Figure 3**. Electrically tunable ion transport of MoS$_2$/SiN nanochannel. (a) Illustration of measurement configuration; (b,c) Ion current of nanochannel under different gate bias and transmembrane voltage (in mV).

Conversely, applying a positive gate voltage resulted in enhanced conductance. These ambipolar characteristics indicate that the nanochannel surface is nearly electrically neutral at zero gate voltage, and the application of either positive or negative gate voltages increases the accumulation of counter-ions, thereby increasing conductance [13,52].

The I–V curves further reveal a switch in ICR behavior depending on the direction of the applied bias. At zero gate voltage, the current is suppressed under positive transmembrane bias, a trend that persists with increasing positive gate voltage. However, when negative gate voltages are applied, the rectification behavior reverses — conductance under negative transmembrane bias decreases relative to that under positive bias. A similar phenomenon was previously reported by Reed's group [19].

Fig 3(c) displayed the gate-dependent ionic current characteristics of our ionic transistor, where an asymmetry can be observed between the response under positive and negative transmembrane potential $V_{SD}$. For negative $V_{SD}$, the ion current quickly saturates at around 300 mV while for positive $V_{SD}$ the $I_{SD}$ increased monotonically. This asymmetric behavior could be attributed to the asymmetric of the $MoS_2$ surface area on cis or trans reservoir, while the surface charge of the smaller area reaches the limited surface charge density faster.

**Protein Translocation**

In a previous study, Dekker's group showed that the 2D material stacks nano-slit (nanochannel) can detect different types of DNA translocation events, thereby introducing such nanochannels as a new tool to probe biopolymer properties [4]. Here we tested the ability of our nanochannel for protein translocations measuring the translocation of BSA. Unlike the evenly distributed negative charge and long strand structure of DNA molecules, protein molecules are normally globular which means the dwell time is quite small which makes detection challenging, and the unevenly charge distribution makes it difficult for the electrical force driving the protein molecule translocating through the nano-pores/channels [53]. While sodium dodecyl sulfate (SDS) has been proved to unfold protein and stabilize this denatured state, the absorbed SDS alongside the unfold protein chain can also introduce uniformed negative charge which can facilitate its electrophoretic translocation through 10 nm solid-state nanopore [54,55].

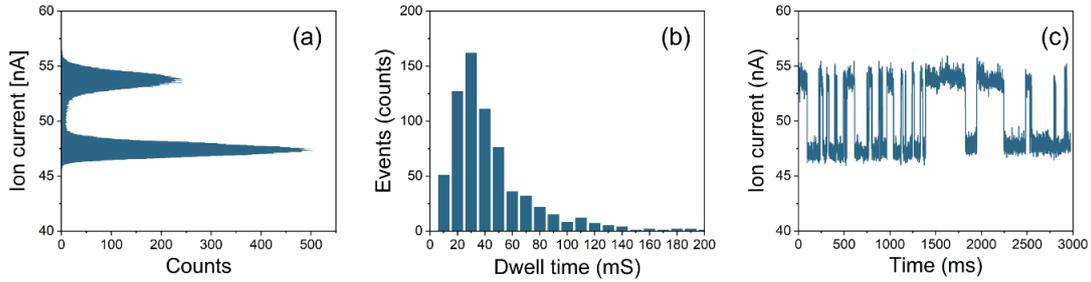

**Figure 4.** Translocation analysis of BSA through the MoS$_2$/SiN nanochannel under an applied voltage of +200 mV. (a) Ion current blockade histogram; (b) Dwell time distribution and (c) Section of translocation trace.

According to the literature, measuring the magnitude of translocation spikes of BSA protein molecules through ~55 nm SiN nanopore with 20 nm thickness, it is possible to estimate the size of the BSA between 7 to 9 nm [56]. After unfolding by SDS the hydrodynamic radii of BSA-SDS were determined to be around 5.9 nm [57]. Here, we used 1 M KCl and 200 μM SDS solution buffered with Tris-EDTA at pH 8 as the electrolyte to measure the translocation of BSA (~8 mg in 4 mL TE buffer solution, ~30 μM). In our MoS$_2$/SiN nanochannel, a voltage of 200 mV is sufficient to initiate the translocation of BSA-SDS chains, which is sufficiently low to prevent damage or delamination of the MoS$_2$ layers [31]. An ionic current trace of the MoS$_2$/SiN nanochannel at a bias voltage of 200 mV is recorded over 5 min [Fig. S5] for translocation events analysis as shown in Fig. 4(a) and Fig. 4(b). A representative 3-second segment highlighting characteristic signal features was selected for detailed analysis and presented in Fig. 4(c). Compared to previously published works of BSA translocation through single-layer MoS$_2$ nanopore or of BSA-SDS translocate through few nanometer size SiN nanopore, we observed a significant enhancement of the translocation dwell time, which could stem from the noncovalent interaction between the disulfide bonds from BSA chain and the S-vacancies from the MoS$_2$ surface [24].

A significant increase in the baseline ionic current was observed in the MoS$_2$/SiN nanochannel after the channel was incubated by low concentration SDS/KCl solution and the high concentration SDS-modified BSA solution was introduced into the nanochannel (Fig. 4c), reaching values exceeding 50 nA compared to the smaller baseline of a few nA measured under standard buffer conditions (Fig. 2). We postulate that this substantial increment arises primarily from the adsorption of excess SDS molecules onto the hydrophobic MoS$_2$ surface. This adsorption likely introduces a high density of additional negative charges, thereby significantly increasing the surface charge density of the channel walls. The enhanced negative surface charge density promotes a stronger attraction of counter-ions (cations) within the electrical double layer. Given the nanoscale confinement of our channel, this could result in pronounced EDL overlaps, thus boosting the surface conductivity and consequently the baseline ionic current. Furthermore, the SDS solution itself inherently contributes a high ionic strength (due to its Na$^+$ counter-ions), which would also elevate the bulk conductivity within the nanochannel.

CONCLUSION

In summary, this study experimentally investigated the ion transport through MoS$_2$/SiN nanochannels with width ~100 nm and depth ~10 nm, demonstrated its ion current tunability through gating electrode and applications on osmotic power and linearized BSA translocation. The ambipolar nanofluidic ion transistor behavior suggests the surface charge tunability of MoS$_2$ flake. Further work can be expected to extend the single nanochannel configuration to multi-channel structures with separate gate electrodes to achieve multiplex tunability which can be used as ionic gate. The large dwell time of linearized BSA translocation indicates the potential of utilizing the interaction between molecules and MoS$_2$ surface for molecule sensing. This work may facilitate future ion fluidic studies along those applications.

ASSOCIATED CONTENT

AUTHOR INFORMATION

**Corresponding Author**

*denis.garoli@unimore.it

**Author Contributions**

SW conceived the system, performed the fabrication and characterization; AD supported with the sample characterization and with supervision; MT supported for the osmotic power experiments, GL and AS supported with the sample design and fabrication, LI, AG and RK supported the experiment discussion, DG supervised the work.

**Funding Sources**

ACKNOWLEDGMENT

The authors thank the European Union under the HORIZON-MSCADN-2022: DYNAMO, grant Agreement 101072818.

REFERENCES

[1] A. Keerthi, S. Goutham, Y. You, P. Iamprasertkun, R. A. W. Dryfe, A. K. Geim, and B. Radha, Water friction in nanofluidic channels made from two-dimensional crystals, Nat Commun **12**, 3092

(2021).

[2] A. Noy and S. B. Darling, Nanofluidic computing makes a splash, Science **379**, 143 (2023).

[3] Y. Cui, H. Guo, X. Yan, W. Zhou, Q. Ye, C. Ying, Z. Liu, and J. Tian, Nonlinear and Anisotropic Ion Transport in Black Phosphorus Nanochannels, Nano Lett. **23**, 5886 (2023).

[4] W. Yang, B. Radha, A. Choudhary, Y. You, G. Mettela, A. K. Geim, A. Aksimentiev, A. Keerthi, and C. Dekker, Translocation of DNA through Ultrathin Nanoslits, Advanced Materials **33**, 2007682 (2021).

[5] W. Xin, H. Ling, Y. Cui, Y. Qian, X.-Y. Kong, L. Jiang, and L. Wen, Tunable Ion Transport in Two-Dimensional Nanofluidic Channels, J. Phys. Chem. Lett. **14**, 627 (2023).

[6] L. Xie, J. Tang, R. Qin, Q. Zhang, J. Liu, Y. Jin, and H. Wang, Surface Charge Modification on 2D Nanofluidic Membrane for Regulating Ion Transport, Advanced Functional Materials **33**, 2208959 (2023).

[7] K. Liu, J. Feng, A. Kis, and A. Radenovic, Atomically Thin Molybdenum Disulfide Nanopores with High Sensitivity for DNA Translocation, ACS Nano **8**, 2504 (2014).

[8] P. Wei, M. Kansari, and M. Fyta, Graphene or MoS 2 nanopores: pore adhesion and protein linearization, (2025).

[9] C. Gu, Z. Yu, X. Li, X. Zhu, Z. Cao, Z. Ye, C. Jin, and Y. Liu, Experimental study of protein translocation through MoS2 nanopores, Applied Physics Letters **115**, 223702 (2019).

[10] S.-C. Chen, C.-Y. Lin, T.-L. Cheng, and W.-L. Tseng, 6-Mercaptopurine-Induced Fluorescence Quenching of Monolayer MoS2 Nanodots: Applications to Glutathione Sensing, Cellular Imaging, and Glutathione-Stimulated Drug Delivery, Advanced Functional Materials **27**, 1702452 (2017).

[11] M. Tsutsui, W.-L. Hsu, D. Garoli, I. W. Leong, K. Yokota, H. Daiguji, and T. Kawai, Gate-All-

Around Nanopore Osmotic Power Generators, ACS Nano **18**, 15046 (2024).

[12] T. Mei, W. Liu, G. Xu, Y. Chen, M. Wu, L. Wang, and K. Xiao, Ionic Transistors, ACS Nano acsnano.3c06190 (2024).

[13] W. Liu et al., Bioinspired carbon nanotube–based nanofluidic ionic transistor with ultrahigh switching capabilities for logic circuits, Sci. Adv. **10**, eadj7867 (2024).

[14] P. Liu, X.-Y. Kong, L. Jiang, and L. Wen, Ion transport in nanofluidics under external fields, Chem. Soc. Rev. **53**, 2972 (2024).

[15] H. Hong, X. Lei, J. Wei, Y. Zhang, Y. Zhang, J. Sun, G. Zhang, P. M. Sarro, and Z. Liu, Rectification in Ionic Field Effect Transistors Based on Single Crystal Silicon Nanopore, Adv Elect Materials 2300782 (2024).

[16] C.-Y. Lin, E. Turker Acar, J. W. Polster, K. Lin, J.-P. Hsu, and Z. S. Siwy, Modulation of Charge Density and Charge Polarity of Nanopore Wall by Salt Gradient and Voltage, ACS Nano **13**, 9868 (2019).

[17] R. Ren, Y. Zhang, B. P. Nadappuram, B. Akpinar, D. Klenerman, A. P. Ivanov, J. B. Edel, and Y. Korchev, Nanopore extended field-effect transistor for selective single-molecule biosensing, Nat Commun **8**, 586 (2017).

[18] N. Di Fiori, A. Squires, D. Bar, T. Gilboa, T. D. Moustakas, and A. Meller, Optoelectronic control of surface charge and translocation dynamics in solid-state nanopores, Nature Nanotech **8**, 946 (2013).

[19] W. Guan, R. Fan, and M. A. Reed, Field-effect reconfigurable nanofluidic ionic diodes, Nat Commun **2**, 506 (2011).

[20] S.-W. Nam, M. J. Rooks, K.-B. Kim, and S. M. Rossnagel, Ionic Field Effect Transistors with

Sub-10 nm Multiple Nanopores, Nano Lett. **9**, 2044 (2009).

[21] Y. Xue, Y. Xia, S. Yang, Y. Alsaid, K. Y. Fong, Y. Wang, and X. Zhang, Atomic-scale ion transistor with ultrahigh diffusivity, Science **372**, 501 (2021).

[22] L. Yang et al., Chloride Molecular Doping Technique on 2D Materials: WS2 and MoS2, Nano Lett. **14**, 6275 (2014).

[23] Y. Li, C.-Y. Xu, P. Hu, and L. Zhen, Carrier Control of MoS2 Nanoflakes by Functional Self-Assembled Monolayers, ACS Nano **7**, 7795 (2013).

[24] X. Chen, N. C. Berner, C. Backes, G. S. Duesberg, and A. R. McDonald, Functionalization of Two-Dimensional MoS2: On the Reaction Between MoS2 and Organic Thiols, Angewandte Chemie **128**, 5897 (2016).

[25] S. Kwon, M. H. Kwon, J. Song, E. Kim, Y. Kim, B. R. Kim, J. K. Hyun, S. W. Lee, and D.-W. Kim, Light-Induced Surface Potential Modification in MoS2 Monolayers on Au Nanostripe Arrays, Sci Rep **9**, 14434 (2019).

[26] C. Nie, B. Zhang, Y. Gao, M. Yin, X. Yi, C. Zhao, Y. Zhang, L. Luo, and S. Wang, Thickness-Dependent Enhancement of Electronic Mobility of MoS2 Transistors via Surface Functionalization, J. Phys. Chem. C **124**, 16943 (2020).

[27] C. J. McClellan, E. Yalon, K. K. H. Smithe, S. V. Suryavanshi, and E. Pop, High Current Density in Monolayer MoS2 Doped by AlOx, ACS Nano **15**, 1587 (2021).

[28] Y. He, M. Tsutsui, C. Fan, M. Taniguchi, and T. Kawai, Controlling DNA Translocation through Gate Modulation of Nanopore Wall Surface Charges, ACS Nano **5**, 5509 (2011).

[29] M. Graf, M. Lihter, M. Thakur, V. Georgiou, J. Topolancik, B. R. Ilic, K. Liu, J. Feng, Y. Astier, and A. Radenovic, Fabrication and practical applications of molybdenum disulfide nanopores, Nat

Protoc **14**, 4 (2019).

[30] V. Tabard-Cossa, D. Trivedi, M. Wiggin, N. N. Jetha, and A. Marziali, Noise analysis and reduction in solid-state nanopores, Nanotechnology **18**, 305505 (2007).

[31] B. Radha et al., Molecular transport through capillaries made with atomic-scale precision, Nature **538**, 222 (2016).

[32] M. A. Alibakhshi, X. Kang, D. Clymer, Z. Zhang, A. Vargas, V. Meunier, and M. Wanunu, Scaled-Up Synthesis of Freestanding Molybdenum Disulfide Membranes for Nanopore Sensing, Advanced Materials **35**, 2207089 (2023).

[33] Y. Huang et al., Universal mechanical exfoliation of large-area 2D crystals, Nat Commun **11**, 2453 (2020).

[34] M. D. Siao, W. C. Shen, R. S. Chen, Z. W. Chang, M. C. Shih, Y. P. Chiu, and C.-M. Cheng, Two-dimensional electronic transport and surface electron accumulation in MoS2, Nat Commun **9**, 1442 (2018).

[35] D. Mosconi et al., Site-Selective Integration of MoS$_2$ Flakes on Nanopores by Means of Electrophoretic Deposition, Acs Omega **4**, 9294 (2019).

[36] B. Yang, B. Bhujel, D. G. Chica, E. J. Telford, X. Roy, F. Ibrahim, M. Chshiev, M. Cosset-Chéneau, and B. J. van Wees, Electrostatically controlled spin polarization in Graphene-CrSBr magnetic proximity heterostructures, Nat Commun **15**, 4459 (2024).

[37] P. J. Zomer, M. H. D. Guimarães, J. C. Brant, N. Tombros, and B. J. Van Wees, Fast pick up technique for high quality heterostructures of bilayer graphene and hexagonal boron nitride, Applied Physics Letters **105**, 013101 (2014).

[38] D. G. Purdie, N. M. Pugno, T. Taniguchi, K. Watanabe, A. C. Ferrari, and A. Lombardo, Cleaning

interfaces in layered materials heterostructures, Nat Commun **9**, 5387 (2018).

[39] J. Feng, K. Liu, M. Graf, D. Dumcenco, A. Kis, M. Di Ventra, and A. Radenovic, Observation of ionic Coulomb blockade in nanopores, Nature Mater **15**, 8 (2016).

[40] D. Stein, M. Kruithof, and C. Dekker, Surface-Charge-Governed Ion Transport in Nanofluidic Channels, Phys. Rev. Lett. **93**, 035901 (2004).

[41] S. Goutham, R. K. Gogoi, H. Jyothilal, G.-H. Nam, A. Ismail, S. V. Pandey, A. Keerthi, and B. Radha, Electric Field Mediated Unclogging of Angstrom-Scale Channels, Small Methods **n/a**, 2400961 (2024).

[42] I. V. Sabaraya, H. Shin, X. Li, R. Hoq, J. A. C. Incorvia, M. J. Kirisits, and N. B. Saleh, Role of Electrostatics in the Heterogeneous Interaction of Two-Dimensional Engineered MoS2 Nanosheets and Natural Clay Colloids: Influence of pH and Natural Organic Matter, Environ. Sci. Technol. **55**, 919 (2021).

[43] K. Lin, Z. Li, Y. Tao, K. Li, H. Yang, J. Ma, T. Li, J. Sha, and Y. Chen, Surface Charge Density Inside a Silicon Nitride Nanopore, Langmuir **37**, 10521 (2021).

[44] G.-C. Liu, M.-J. Gao, W. Chen, X.-Y. Hu, L.-B. Song, B. Liu, and Y.-D. Zhao, pH-modulated ion-current rectification in a cysteine-functionalized glass nanopipette, Electrochemistry Communications **97**, 6 (2018).

[45] M. Tsutsui, K. Yokota, I. W. Leong, Y. He, and T. Kawai, Protocol for preparation of solid-state multipore osmotic power generators, STAR Protocols **4**, 102227 (2023).

[46] J. Feng, M. Graf, K. Liu, D. Ovchinnikov, D. Dumcenco, M. Heiranian, V. Nandigana, N. R. Aluru, A. Kis, and A. Radenovic, Single-layer MoS2 nanopores as nanopower generators, Nature **536**, 197 (2016).

[47] Sparse multi-nanopore osmotic power generators, Cell Reports Physical Science **3**, 101065 (2022).

[48] S. Liang, F. Xiang, Z. Tang, R. Nouri, X. He, M. Dong, and W. Guan, Noise in nanopore sensors: Sources, models, reduction, and benchmarking, Nanotechnology and Precision Engineering **3**, 9 (2020).

[49] M. Graf, M. Lihter, D. Unuchek, A. Sarathy, J.-P. Leburton, A. Kis, and A. Radenovic, Light-Enhanced Blue Energy Generation Using MoS2 Nanopores, Joule **3**, 1549 (2019).

[50] J.-P. Hsu, T.-C. Su, P.-H. Peng, S.-C. Hsu, M.-J. Zheng, and L.-H. Yeh, Unraveling the Anomalous Surface-Charge-Dependent Osmotic Power Using a Single Funnel-Shaped Nanochannel, ACS Nano **13**, 13374 (2019).

[51] A. Siria, P. Poncharal, A.-L. Biance, R. Fulcrand, X. Blase, S. T. Purcell, and L. Bocquet, Giant osmotic energy conversion measured in a single transmembrane boron nitride nanotube, Nature **494**, 455 (2013).

[52] R. Fan, M. Yue, R. Karnik, A. Majumdar, and P. Yang, Polarity Switching and Transient Responses in Single Nanotube Nanofluidic Transistors, Phys. Rev. Lett. **95**, 086607 (2005).

[53] C. Plesa, S. W. Kowalczyk, R. Zinsmeester, A. Y. Grosberg, Y. Rabin, and C. Dekker, Fast Translocation of Proteins through Solid State Nanopores, Nano Lett. **13**, 658 (2013).

[54] L. Restrepo-Pérez, S. John, A. Aksimentiev, C. Joo, and C. Dekker, SDS-assisted protein transport through solid-state nanopores, Nanoscale **9**, 11685 (2017).

[55] N. Soni, N. Freundlich, S. Ohayon, D. Huttner, and A. Meller, Single-File Translocation Dynamics of SDS-Denatured, Whole Proteins through Sub-5 nm Solid-State Nanopores, Acs Nano **16**, 11405 (2022).

[56] A. Han, G. Schürmann, G. Mondin, R. A. Bitterli, N. G. Hegelbach, N. F. De Rooij, and U. Staufer,

Sensing protein molecules using nanofabricated pores, Applied Physics Letters **88**, 093901 (2006).

[57] A. Valstar, M. Almgren, W. Brown, and M. Vasilescu, The Interaction of Bovine Serum Albumin with Surfactants Studied by Light Scattering, Langmuir **16**, 922 (2000).

SUPPORTING INFORMATION

Fig. S1. The conductance of MoS2/SiN nanochannel within different concentration under -500, 500 and 1000 mV potential.

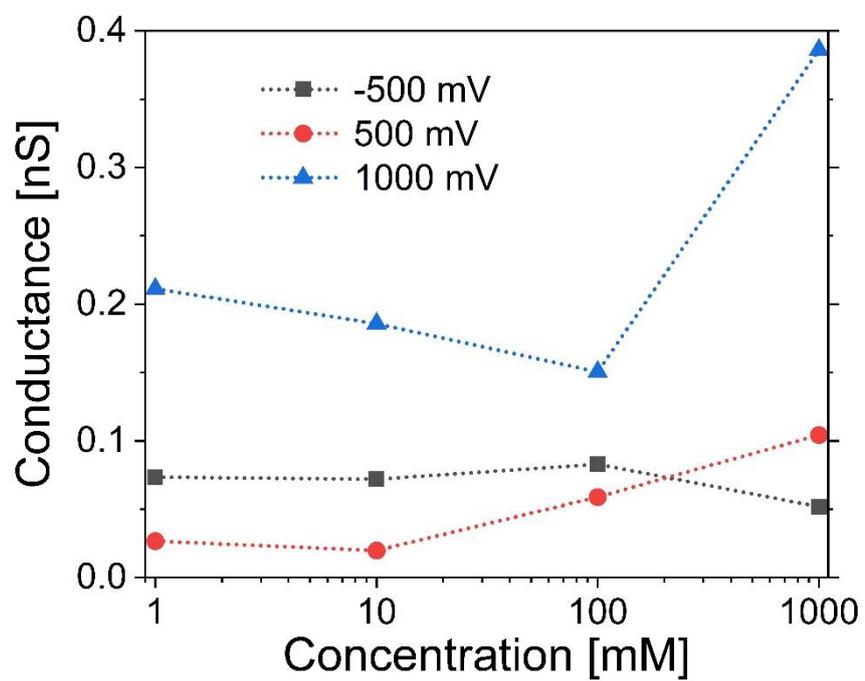

Fig. S2. Ion current traces of low voltage scanning to remove the potential contamination inside the MoS$_2$/SiN channel. (a) 1$^{st}$ to 10$^{th}$ scan; (b) 11$^{th}$ to 268$^{th}$ scan.

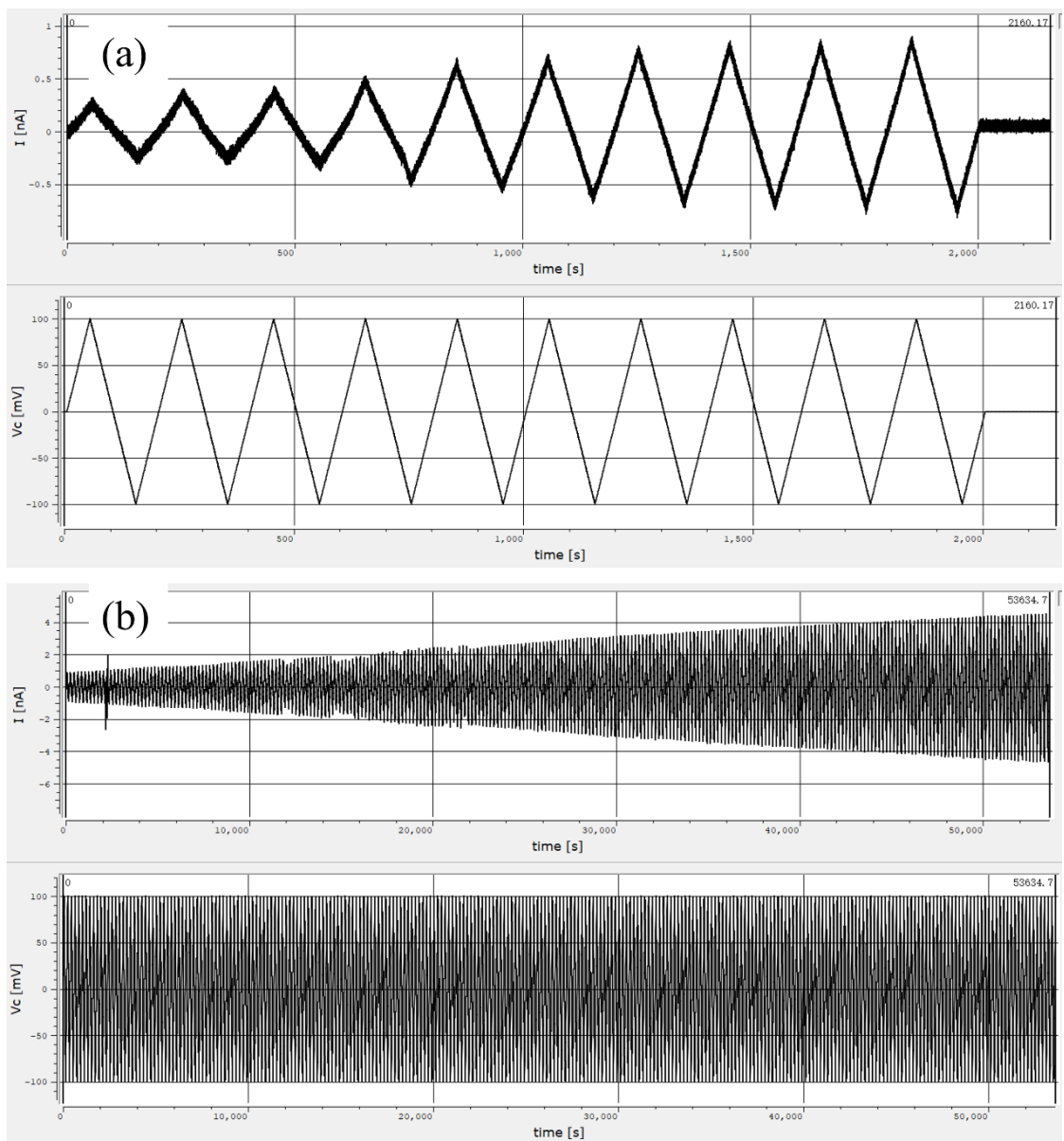

Fig. S3. Power Spectrum Density of nanopore measurement under different electrolyte gradients.

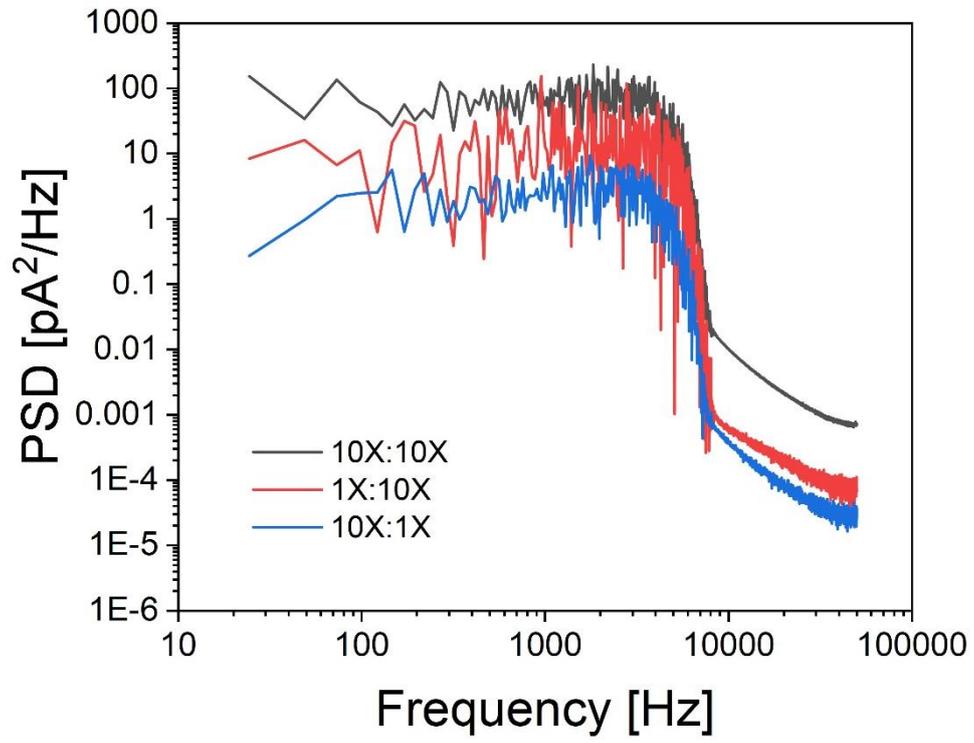

Fig. S4. Scaling of the osmotic power of MoS$_2$/SiN nanochannel.

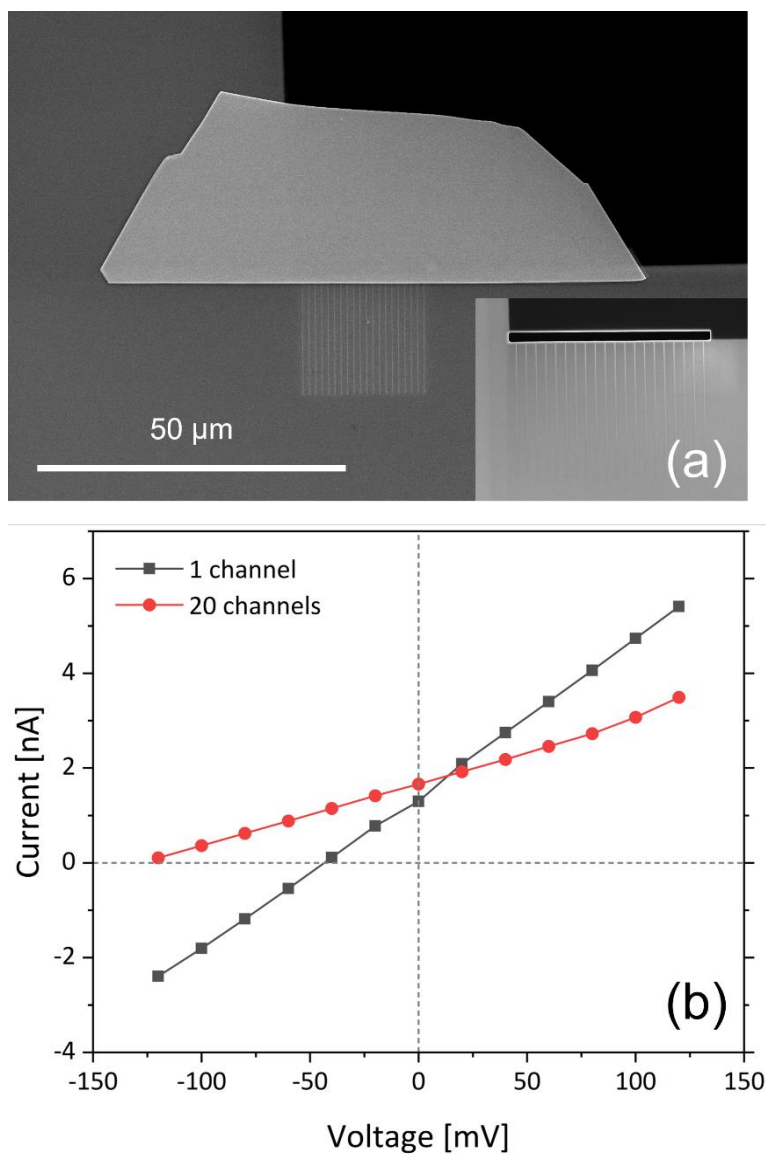

Fig. S5. BSA translocation events during 2 min recording.

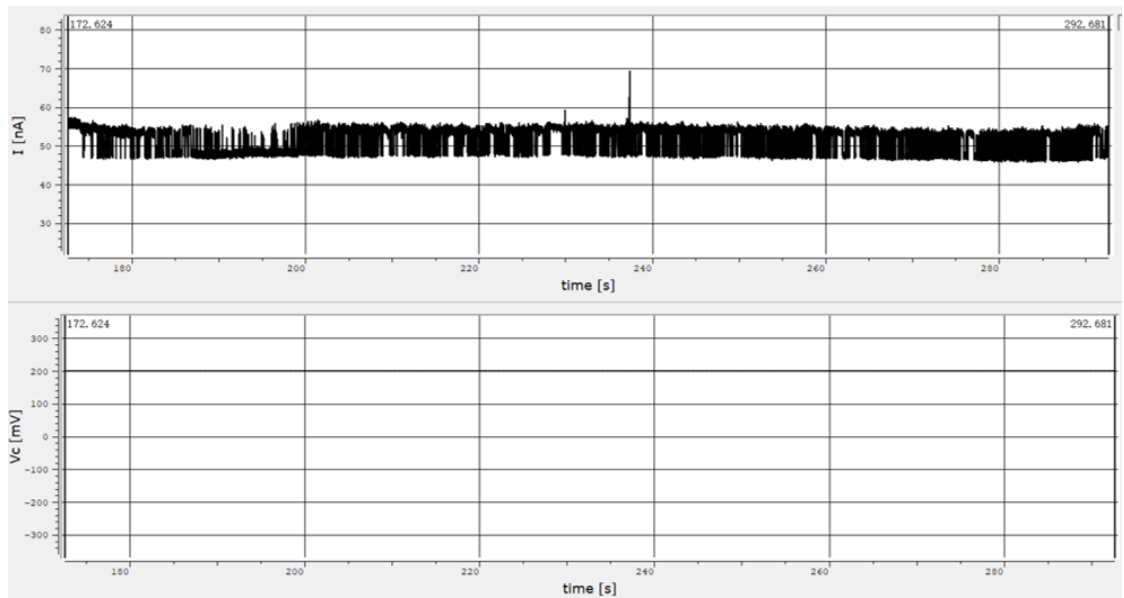